\documentclass[preprint,showpacs,preprintnumbers,amsmath,amssymb]{revtex4}
\usepackage{graphicx}
\usepackage{rotating}
\usepackage{epsf}
\usepackage{dcolumn}
\begin{document}
\preprint{preprint - thermoelectric group, iitp/iitb}
\title{n-type SnSe$_{1-x}$ for Thermoelectric Application}
\author{Tutul Bera$^{1}$, Anup V. Sanchela$^{2}$, C. V. Tomy$^{2}$, Ajay D. Thakur$^{1,\,3}$}
\affiliation{$^{1}$ Department of Physics, School of Basic Sciences, Indian Institute of Technology Patna, Patna 801118, India \\
$^{2}$ Department of Physics, Indian Institute of Technology Bombay, Mumbai 400076, India 
\\
$^{3}$ Center for Energy and Environment, Indian Institute of Technology Patna, Patna 801118, India}
\date{\today}
\begin{abstract}
We report the synthesis of n-type SnSe$_{1-x}$ using a self-sacrificial, facile, solvo-thermal synthesis route. Electrical and thermal transport measurements suggest a low thermal conductivity and a significant thermopower in the temperature range 100 - 400\,K. We also propose the possibility of developing an all SnSe thermoelectric module.
\end{abstract}
\pacs{74.25.Dw, 74.70.Xa}
\maketitle

\section{Introduction}
Thermoelectric (TE) materials have tremendous potential for applications in the areas of energy harvesting, chip cooling, noiseless HVAC (heating-ventilation and air-conditioning) systems for point-of-care medical purpose, etc. Considerable amount of research has went into finding TE materials with a high figure of merit ($zT$), where, $zT = \frac{S^2\sigma}{\kappa}$; S is the Seebeck coefficient, $\sigma$ is the electrical conductivity and $\kappa$ is the thermal conductivity. 
Recent reports suggest SnSe as a promising p-type material for TE applications \cite{ref1, ref2, ref3}. Realization of TE modules based on p-type SnSe requires a suitable n-type counterpart. In fabrication of TE modules, there are critical material compatibility issues in choosing the respective p-type and n-type materials. Whereas repeated thermal cycling leads to degradation in the performance of TE modules, there are additional limitations arising out of manufacturability constraints \cite{ref4, ref5, ref6}. From the perspective of making on-chip integrated TE module, it is desirable to reduce the material complications arising out of different deposition requirements for different materials for the p-type and the n-type materials. Theoretical investigation suggested the possibility of an n-type carrier response in SnSe upon suitable doping \cite{ref7}. These n-type materials were predicted to have a higher $zT$ than their p-type counterparts. In this work, we report the experimental realization of n-type SnSe via introduction of Se site deficiency.

\section{Experimental Details}
Anhydrous tin(II) chloride (SnCl$_2$), selenourea (CH$_4$N$_2$Se) were procured from Sigma Aldrich. Ethylene glycol (EG), Ethanol were of analytical grade and used without further purification. De-ionized water was obtained from a Milli-Q purification system.
In a typical experimental procedure, 2 mmol of tin(II) chloride was dissolved  in 20 ml EG under constant magnetic stirring. Then 2 mmol selenourea was added to the solution. The resulting transparent solution was loaded into Teflon-lined stainless steel autoclave such that one-fourth of the autoclave remain empty. It was heated to 180$^o$C at a uniform rate of 8$^o$/minute and kept for 24 hours. After cooling, the product was washed 4 times by Milli-Q water and and 3 times by ethanol. Finally the product was dried at 60$^o$C in vacuum for further characterization.
Powder X-ray diffraction (PXRD) of the as synthesized samples was performed using Rigaku TTRAX III. Sample morphology was studied via field emission scanning electron microscopy (FESEM) using Hitachi S-4800.  The energy-dispersive X-ray analysis (EDXA) for determining the nominal composition of the sample was also performed using Hitachi S-4800 FESEM. The temperature ($T$) variation of Seebeck coefficient ($S$) and the thermal conductivity ($\kappa$) were measured in a two probe configuration using the thermal transport option (TTO) of the physical property measurements system (PPMS) from Quantum Design Inc. The temperature variation of electrical resistivity ($\rho$) was measured in a four probe configuration using the resistivity option of PPMS.

\section{Results and Discussion}
The powder X-ray diffraction (PXRD) pattern of the as synthesized sample is shown in Fig.~1. All of the diffraction peaks can be indexed to orthorhombic phase with $Cmcm$ space group. The stick diagram from ICSD (Inorganic Crystal Structure Database) of standard orthorhombic SnSe (NO: 01-083-0253 CSD: 50563) is also attached for comparison. Panels (a), (b) and (d) of Fig.~2 show the FESEM images of the as grown sample at various magnifications. It is seen that our sample possesses flower-like hierarchical morphology. The EDXA of our samples suggest a Sn rich (Se poor) composition with a chemical formula SnSe$_{0.96}$ (see Fig.~2 (c) for details).
The stoichiometric SnSe reported in literature was a p-type semiconductor \cite{ref1, ref2, ref3}. The nominal composition of the as synthesized sample indicates a Se deficiency. Variation of $S$ with $T$ is shown in Fig.~3. The negative value of $S$($T$) over the entire measurement range from 100-400\,K suggests the n-type character of the Se deficient sample. This points to the possibility of introducing n-type character by Se deficiency. Such an anion vacancy induced n-type character was seen in certain oxides including MgO, ZnO, In$_2$O$_3$, SnO$_2$, Al$_2$O$_3$, Fe$_2$O$_3$,TiO$_2$, etc \cite{ref8}. Tanaka {\it et.al}  \cite{ref9} looked into the details of the origin on n-type conductivity in these materials using first-principle calculations.  Similar anion vacancies also occur in metal chalcogenide, which was described theoretically in Cu-III-VI$_2$ chalcopyrite systems \cite{ref11, ref12}. Figure~3 also shows the variation of $\kappa$ as a function of $T$. The inset panel of Fig.~3 shows the variation of $\rho$ with $T$. The small values for $\kappa$ along with a small $\rho$ and a reasonably large $S$ points to the utility of n-type SnSe$_{1-x}$ as a promising TE material as predicted theoretically by Kutorasinski {\it et al} \cite{ref7}. In Fig.~4 we show the variation of power-factor as a function of temperature within a temperature range 100-400~K.\\

A wide variety of technique has been developed in recent past to make high efficiency TE modules. There is a  significant amount of difficulty associated with fabricating an on-chip TE module using different materials for the p-type and n-type legs. The related challenges can be significantly reduced if the two legs can be based on the same material system. The efficiency of the TE device is depends on material compatibility with both p type and n type in the basis of their different electrical and thermal properties. In this regard SnSe presents itself a potential material for TE applications, as it can be easily converted from a p-type material to an n-type material by suitable introduction of Se deficiency. The converse is also feasible employing suitable selenization process. We propose a typical fabrication process for making a TE module with SnSe/SnSe$_{1-x}$ as p-type/n-type legs as shown schematically in Fig.~5 and Fig.~6.

In conclusion, we have successfully synthesized n-type SnSe$_{1-x}$ via introduction of Se deficiency using a self-sacrificial, facile, solvo-thermal synthesis route. We also propose a strategy for making an on-chip TE module using SnSe/SnSe$_{1-x}$ as p-type/n-type legs.

CVT would like to acknowledge the Department of Science and Technology for partial support through the project IR/S2/PU-10/2006.  ADT acknowledges partial support from the Center for Energy and Environment, Indian Institute of Technology, Patna.

\newpage		

\begin{figure} 
\includegraphics[height=13.0 cm]{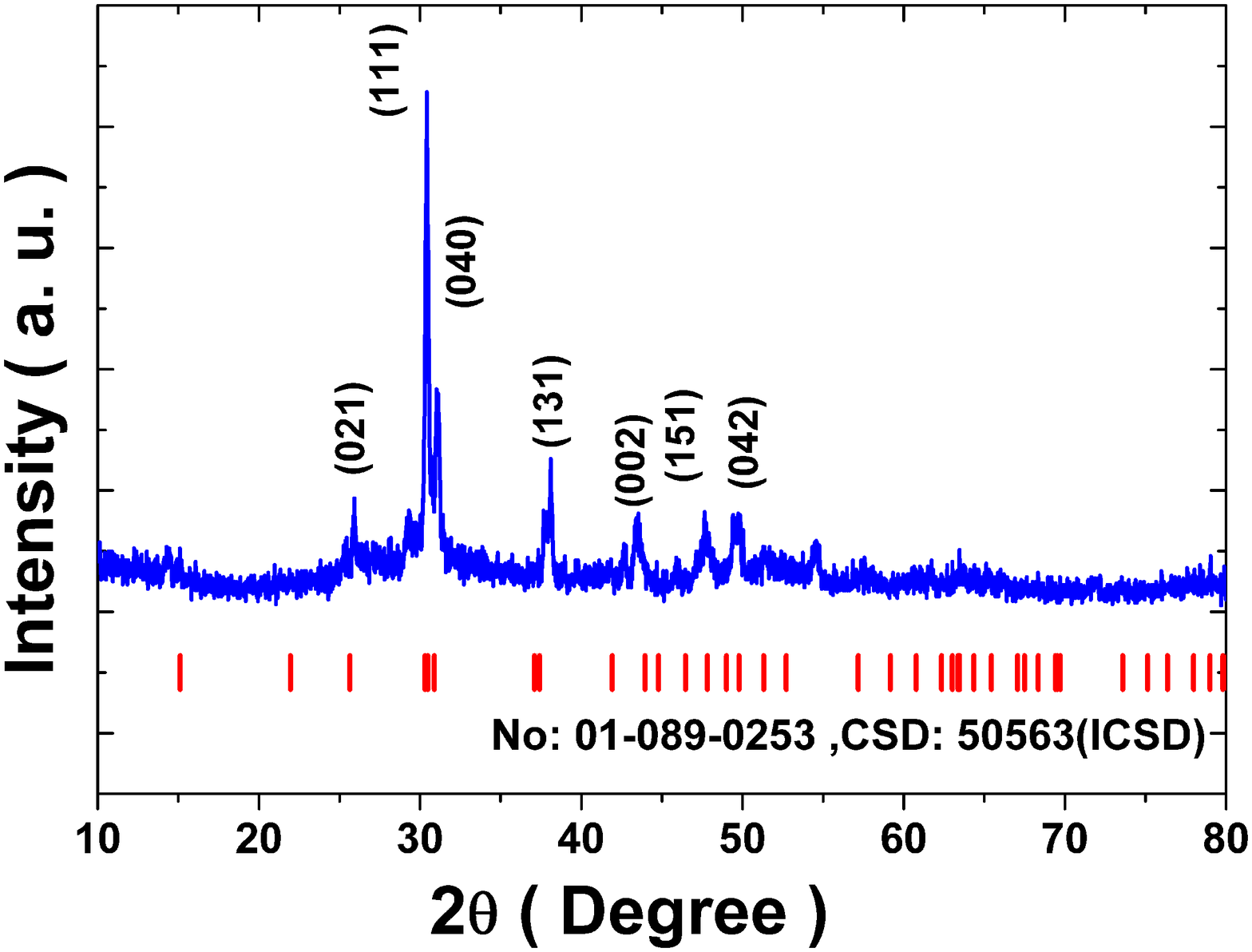}
\caption{(Color online) Powder X-ray Diffraction ( XRD) pattern for the as grown sample of SnSe$_{1-x}$. The stick diagram from ICSD (Inorganic Crystal Structure Database) for SnSe$_{1-x}$ (NO: 01-083-0253 CSD: 50563) is also shown and the peaks have been indexed accordingly.}
\end{figure}

\begin{figure} 
\includegraphics[height=12.0 cm]{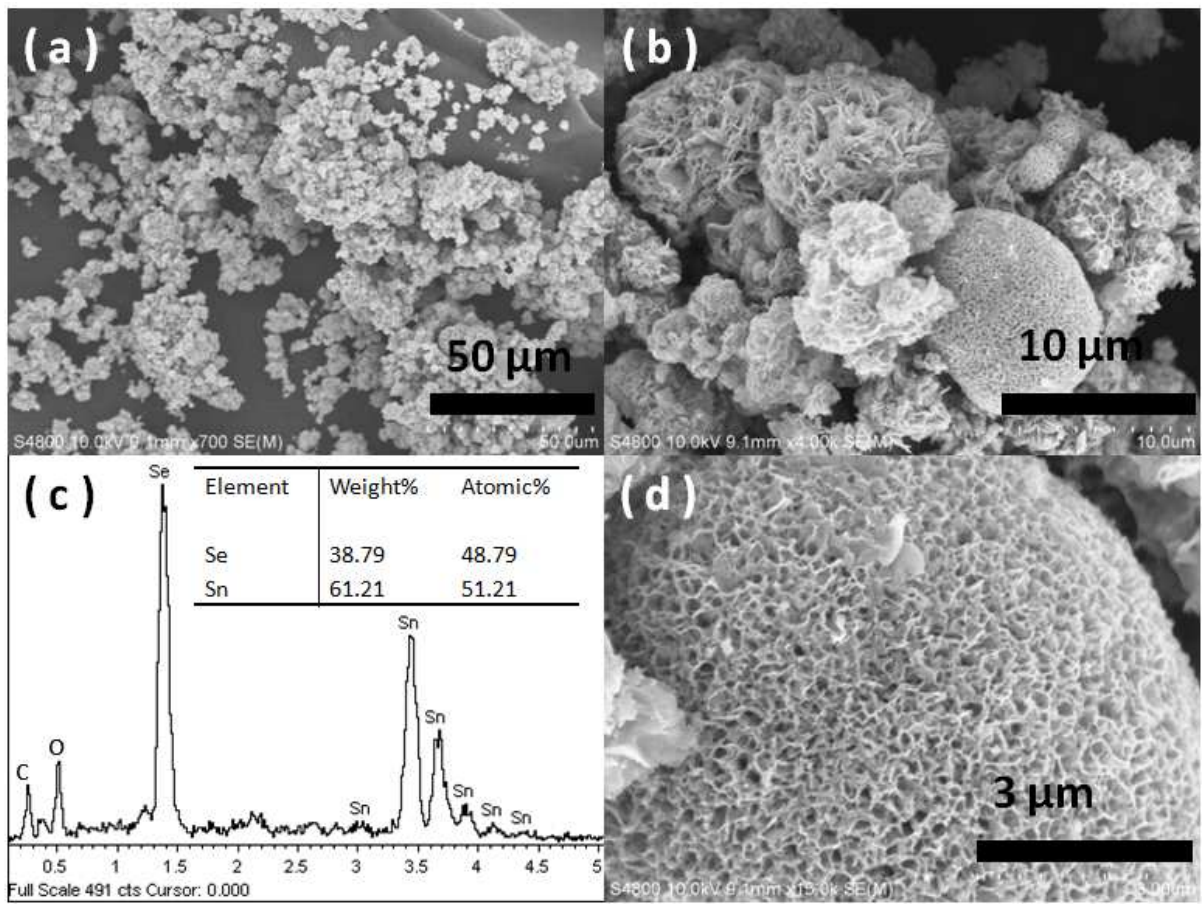}
\caption{(Color online) (a), (b), (d): SEM image at various magnification of the as grown SnSe$_{1-x}$ sample. (c): Energy Dispersive X-ray analysis spectrum of the as grown SnSe$_{1-x}$ sample. The weight$\%$ and atomic$\%$ are shown in the inset.}
\end{figure}

\begin{figure} 
\includegraphics[height=12.0 cm]{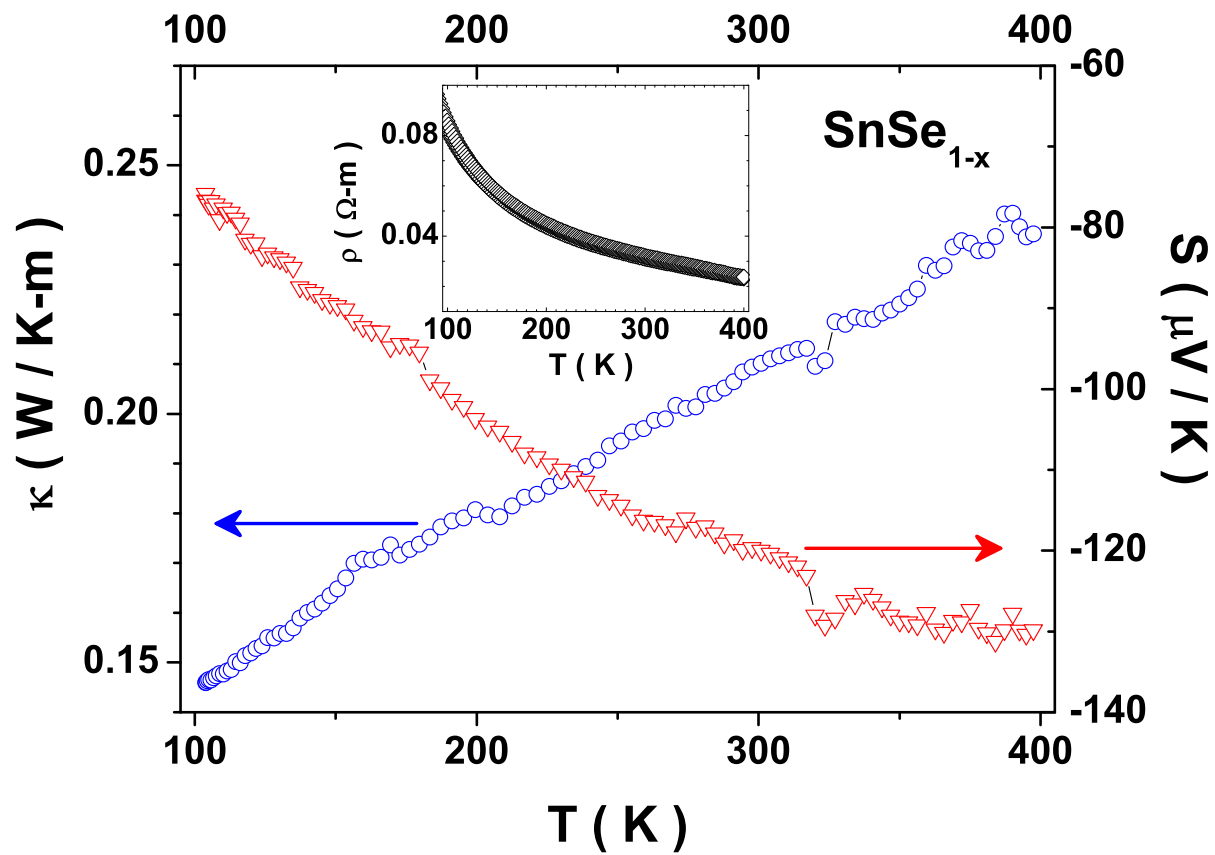}
\caption{(Color online) Variation of $S$ and $\kappa$ as a function of $T$. Inset panel shows the variation of $\rho$ as a function of $T$.}
\end{figure}

\begin{figure} 
\includegraphics[height=13.0 cm]{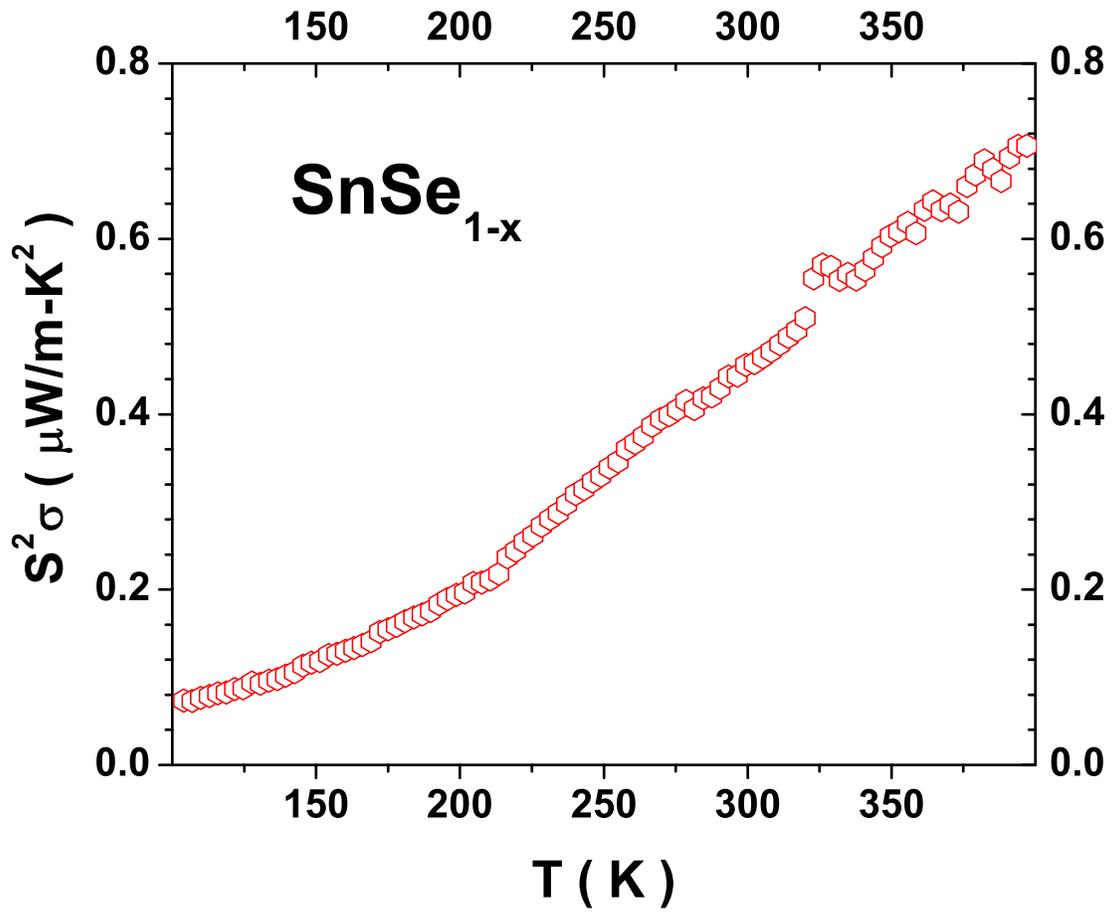}
\caption{(Color online) Variation of power factor ($S^2\sigma$) as a function of $T$.}
\end{figure}

\begin{figure} 
\includegraphics[height=9.0 cm]{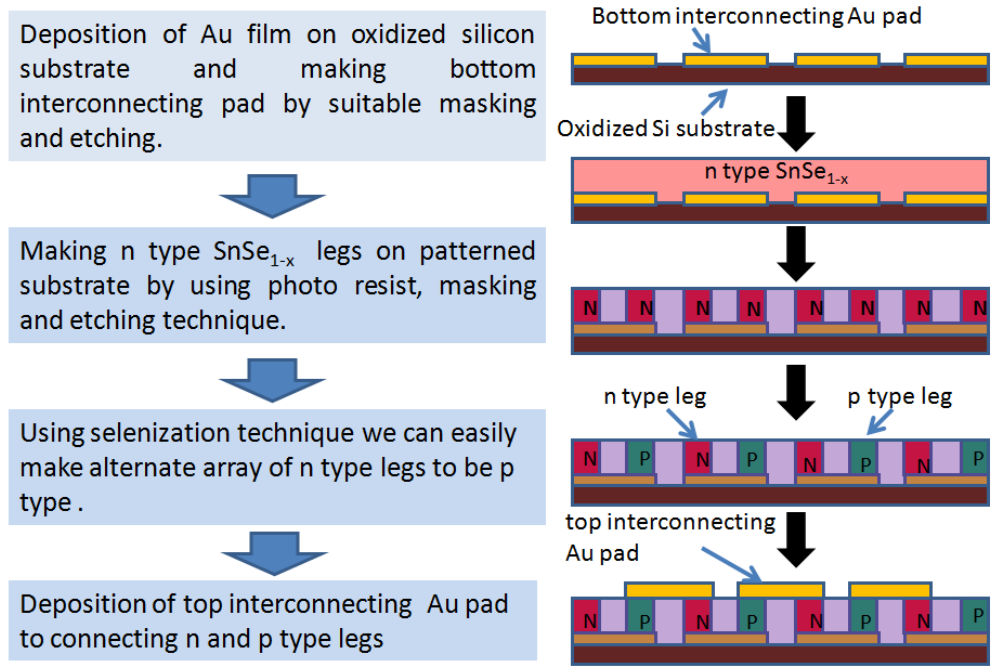}
\caption{(Color online) Schematic of the fabrication process for the proposed SnSe based TE module.}
\end{figure}

\begin{figure} 
\includegraphics[height=12.0 cm]{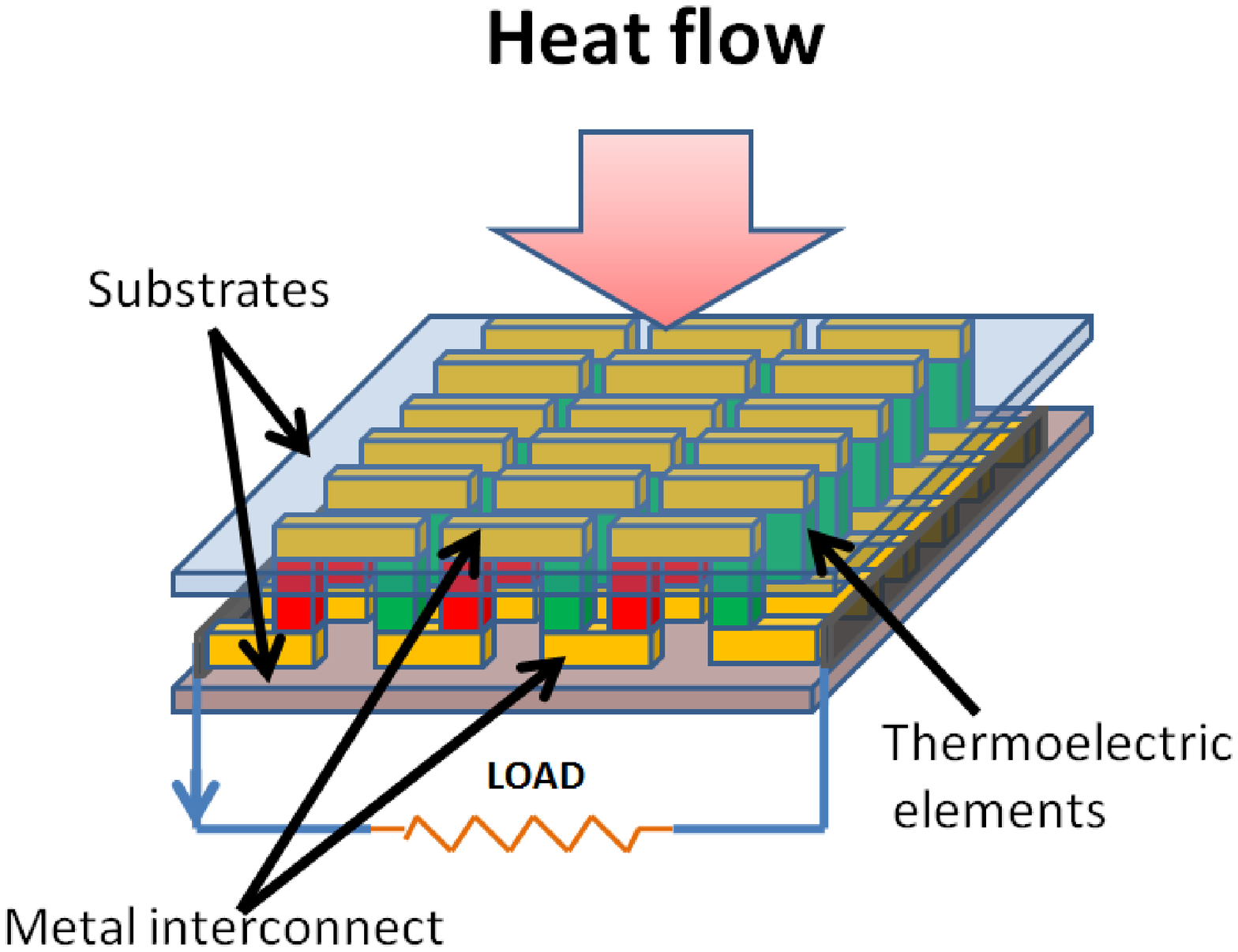}
\caption{(Color online) Schematic of a typical TE device. Both n-type and p-type thermoelectric elements are sandwiched between two high-thermal-conductivity substrates.}
\end{figure}


\begin{references}

\bibitem{ref1}
Li-Dong Zhao, Shih-Han Lo, Yongsheng Zhang, Hui Sun, Gangjian Tan, Ctirad Uher, C. Wolverton, Vinayak P. Dravid and Mercouri G. Kanatzidis, Nature 508, 373 (2014).
\bibitem{ref2}
Jes\'{u}s Carrete, Natalio Mingo, and Stefano Curtarolo, Appl. Phys. Lett. 105, 101907 (2014).
\bibitem{ref3}
S. Sassi, C. Candolfi, J.-B. Vaney, V. Ohorodniichuk, P. Masschelein, A. Dauscher, and B. Lenoir, 104, 212105 (2014).
\bibitem{ref4}
S. Leblanc, Sustainable Materials and Technologies, 1, 26 (2014).
\bibitem{ref5}
Nancy Yang and Alfredo Morales, Sandia Reports, SAND2009-0758, February 2009.
\bibitem{ref6}
Pham Hoang Ngan, Dennis Vabjorn Christensen, Gerald Jeffrey Snyder, Le Thanh Hung, Soren Linderoth, Ngo Van Nong and Nini Pryds, Phys. Status. Solidi A 211, 9 (2014).
\bibitem{ref7}
K. Kutorasinski, B. Wiendlocha, S. Kaprzyk, and J. Tobola, Phys. Rev. B 91, 2015201 (2015).
\bibitem{ref8}
G. D. Mahan, J. Appl. Phys. 54, 3825 (1983).
\bibitem{ref9}
I. Tanaka, K.  Tatsumi, M. Nakano, and H. Adachi, J. Am. Ceram. Soc.,85, 68 –74 (2002).
\bibitem{ref10}
S. Lany and A. Zungar, Phys. Rev. B 72, 035215 (2005).
\bibitem{ref11}
C. Persson,Y. J.  Zhao, S. Lany and A. Zungar, Phys. Rev. B 72, 035211 (2005).

\end{references}
\end{document}